\documentclass[a4paper,twocolumn]{revtex4-1}
\usepackage{amsmath,amssymb}
\usepackage[pdftex]{graphicx}
\usepackage{here}
\usepackage{comment}
\usepackage{color}
\newcommand{\setmid}{\mathrel{}\middle|\mathrel{}}

\def\TDF{table displacement function }
\def\Okninskiquad{the piecewise quadratic case }
\def\Okninskicubic{the piecewise cubic case }

\begin{document}
\title{Local bifurcation structure of a bouncing ball system with a piecewise polynomial function for table displacement}
\author{Yudai Okishio}
\author{Hiroaki Ito}
\author{Hiroyuki Kitahata}
\email{kitahata@chiba-u.jp}
\affiliation{Department of Physics, Chiba Univeristy, Yayoi-cho 1-33, Inage-ku, Chiba 263-8522, Japan}
\date{\today}

\title{Local bifurcation structure of a bouncing ball system with a piecewise polynomial function for table displacement}  

\begin{abstract}
    The system in which a small rigid ball is bouncing repeatedly on a heavy flat table vibrating vertically, so-called the bouncing ball system, has been widely studied.
    Under the assumption that the table is vibrating with a piecewise polynomial function of time, the bifurcation diagram changes qualitatively depending on the order of the polynomial function.
    We elucidate the mechanism of the difference in the bifurcation diagrams by focusing on the two-period solution.
    In addition, we derive the approximate curve of the branch close to the period-doubling bifurcation point in the case of the piecewise cubic function of time for the table vibration.
    We also performed numerical calculation, and we demonstrate that the approximations well reproduce the numerical results.
\end{abstract}

\maketitle 

\section{Introduction}

The bouncing ball system, which consists of a ball bouncing on a heavy flat table vibrating vertically under the gravity, 
is simple yet shows rich dynamical behaviors, such as chattering, periodic motions, and chaos. 
Therefore, many physicists have been attracted to the system since the first systematic study by Holmes \cite{Holmes1982} as a derivative of Fermi's proposal \cite{Fermi1949}.
Experimental studies have also been performed, which have reported that periodic motions and chaos appears depending on the vibration frequency \cite{Tufillaro1986chaotic, Kini2006}.
In relation to more realistic systems, such as granular systems, the horizontal motion of a ball has also been studied using a model with a non-flat table shape \cite{Bae2004, Mcbennett2016, Halev2018}
and a model with a dumbbell-shaped anisotropic object instead of a ball \cite{Dorbolo2005}.

Analytical handling of these systems is difficult because it is necessary to solve a nonlinear equation to find the next collision time.
Further, vibration of the table was typically assumed to be a sinusoidal function of time for the relevance to physical phenomena.
This assumption makes it more difficult to solve the equation since the equation is transcendental in this case.
In order to facilitate analytical handling, 
it is assumed that the maximum height of a bouncing ball is much larger than the amplitude of the table vibration in previous studies \cite{Holmes1982, Mehta1990, Fukano2002, Barroso2009}.
On the other hand, low-order polynomials have been used as alternatives to the sinusoidal function for the \TDF in the last decade.
The equation can always be solved under this assumption, and analytical handling may become easier.
Such studies with low-order polynomials include the piecewise linear case \cite{Okninski2009, Okninski2011},
\Okninskiquad \cite{Okninski2012}, 
and \Okninskicubic \cite{Okninski2013, Li2017, Qiu2017}.
The periodic motions become unstable through the period-doubling bifurcation for the sinusoidal case \cite{Luo1996}.
This is also true for \Okninskicubic \cite{Okninski2013}.
However, the bifurcation diagram in \Okninskiquad is qualitatively different from those in the sinusoidal and the piecewise cubic cases in that 
chaos appears immediately after the destabilization of the fixed point instead of the period-doubling bifurcation \cite{piecewiselinear}.
Therefore, we aim to clarify the mechanism of the qualitative difference of the bifurcation structure between \Okninskiquad and the piecewise cubic case.

In this paper, we primarily discuss the bouncing ball system with a periodic piecewise polynomial function for the table displacement.
We analyze the linear stability of the one-period solution and prove that the bifurcation to a two-period solution does not exist when the \TDF is a piecewise quadratic function.
We analytically derive an approximation curve of the branch close to the period-doubling bifurcation point for the piecewise cubic function.
Furthermore, we propose a function smoothly connecting the piecewise quadratic model \cite{Okninski2012} and the piecewise cubic model \cite{Okninski2013}.
We also performed the numerical calculation for the motion of the ball, and we confirm that the numerical result matches the analytical result.

\section{Model}
We consider a system where a small rigid ball is bouncing on a sufficiently heavy flat table vibrating vertically under the gravitational acceleration $g$ as shown in Fig. 1.
The vertical position of the table follows a given function of time $f(t)$.
We assume that the position $f(t)$ can be written as $f(t)=\alpha F(t)$, where $\alpha ~(>0)$ corresponds to the amplitude of the table vibration.
If we choose the function appropriately, for example $f(t)=\alpha\sin 2\pi t$, the ball repeatedly bounces on the table.
In this situation, $z$, $u$, and $v^{}$ denote the position, the incident velocity, and the reflection velocity of the ball, respectively.
The variables concerning the $i$-th collision are represented with an index $i$.
For simplicity, we do not consider friction or air resistance.
From Newton's law, the position of a ball between the $i$-th collision and the $(i+1)$-th collision, $z_i(t)$, is given by
\begin{align}
    z_i(t) &= -\frac{g}{2}(t-t_i)^2 + v_i(t-t_i)+f(t_i). \label{dob}
\end{align}
Therefore, if we have the sequence of $t_i$ and $v_i$, the dynamics of the ball can be known \cite{completeinelastic}.
$t_{i+1}$ is obtained by solving the nonlinear equation $z_i(t)=f(t)$.
For $v_{i+1}$, Eq. (\ref{dob}) yields
\begin{align}
    u_{i+1} &= v_i-g(t_{i+1}-t_i) ,
\end{align}
and considering that the table is heavy, we have
\begin{align}
    v_{i}^{}-\dot{f}(t_{i}) = -r(u_i-\dot{f}(t_i)) ,
\end{align}
where $r\in (0, 1)$ is the coefficient of restitution and dot represents differential by time $t$.
By setting $\tilde{t}=t/T$, $\tilde{v}=v/(gT)$, $\tilde{z}=z/(gT^2)$
with a characteristic time scale $T$, which is typically the period of $f(t)$,
we obtain the map for $(\tilde{t}_i, \tilde{v}_i)$ in a nondimensional form
\begin{align}
    & \left[
	\begin{array}{c}
        \tilde{t}_{i+1} \\
        \tilde{v}_{i+1}
	\end{array}
    \right] 
    =
    \left[
	\begin{array}{c}
        \mathrm{min}\left\{\tilde{t}\in (\tilde{t}_i, \infty ) \setmid \tilde{z}_i(\tilde{t})=f(\tilde{t})\right\} \\
        (1+r)\dot{f}(\tilde{t}_{i+1}) - r \{\tilde{v}_i-(\tilde{t}_{i+1}-\tilde{t}_i)\}
	\end{array}
    \right] . \label{map}
\end{align}
From here, tildes for nondimensional variables are to be omitted, and we consider the map in Eq. (\ref{map}) since the dynamics of the ball is governed only by this.
We assume that the vertical position of the table $f(t)$ is $\mathrm{C}^1$-class and periodic with a period $T$.
\begin{figure}[H]
  \centering
  \includegraphics{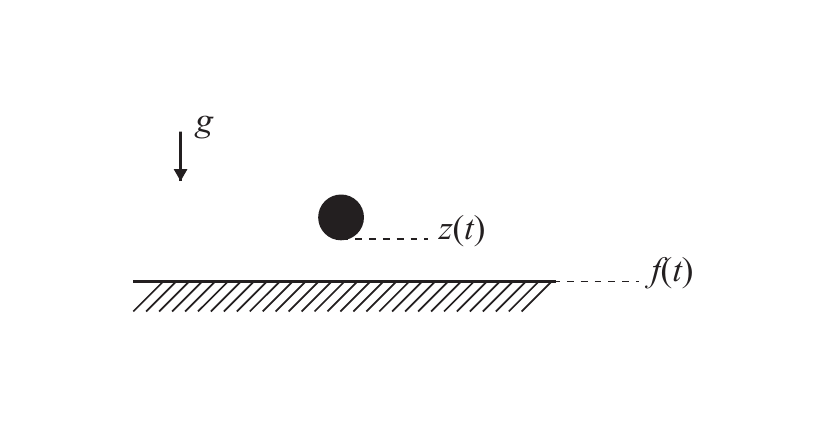}
  \caption{Schematic illustration of the bouncing ball system.}
  \label{fig1}
\end{figure}

In summary, the dynamics of the ball is described by the two state variables: the collision time $t_i$ and the reflection velocity $v_i$, 
with the two control parameters: the coefficient of restitution $r$ and the amplitude of the table vibration $\alpha $.
We mainly discuss the bifurcation structure by varying $\alpha $.

\section{Analytical result}

\subsection{Definition of $(n, k)$-solution}
We formulate periodic solutions by considering the phase of $f(t)$ in this subsection.
Equation (\ref{map}) does not have the periodic solutions in terms of the map for $(t_i, v_i)$ because $t_i < t_{i+1}$ must always be satisfied.
Since time $t$ is normalized by the period of $f(t)$,
we can introduce the fractional part of $t$ as the phase
\begin{align}
    \hat{t} \equiv t - \lfloor t \rfloor ,
\end{align}
where $\lfloor \cdot \rfloor$ is the floor function.
The solution which satisfies the condition 
\begin{gather}
    \left[
	\begin{array}{c}
        t_{i+n} \\
        v_{i+n} 
	\end{array}
    \right]
    =
    \left[
	\begin{array}{c}
        t_i \\
        v_i
	\end{array}
    \right] + 
    \left[
	\begin{array}{c}
        k \\
        0
	\end{array}
    \right]
\end{gather}
such that
\begin{gather}
    \left[
	\begin{array}{c}
        t_{i+j} \\
        v_{i+j} 
	\end{array}
    \right]
    \neq
    \left[
	\begin{array}{c}
        t_i \\
        v_i
	\end{array}
    \right] + 
    \left[
	\begin{array}{c}
        l \\
        0
	\end{array}
    \right]
\end{gather}
with $j = 1, 2, \dots , n-1$ and $l = 1, 2, \dots , k-1$,
is called $(n, k)$-solution, that is to say, the ball exhibits $n$ bounces during $k$ table vibrations $(n, k\in \mathbb{N})$.
The solution fulfills
\begin{gather}
    \left[
	\begin{array}{c}
        \hat{t}_{i+n} \\
        v_{i+n} 
	\end{array}
    \right]
    =
    \left[
	\begin{array}{c}
        \hat{t}_i \\
        v_i
	\end{array}
    \right]
\end{gather}
such that
\begin{gather}
    \left[
	\begin{array}{c}
        \hat{t}_{i+j} \\
        v_{i+j} 
	\end{array}
    \right]
    \neq
    \left[
	\begin{array}{c}
        \hat{t}_i \\
        v_i
	\end{array}
    \right]
    \label{periodic}
\end{gather}
with $j = 1, 2, \dots , n-1$.
This means the solution is $n$-period with respect to the phase of $f(t)$.

\subsection{$(1$, $k)$-solution}

First, we consider the $(1, k)$-solution to know the point at which the bifurcation occurs.
We set this solution as follows
\begin{align}
    \left[
	\begin{array}{c}
        \hat{t}_0 \\
        v_0
	\end{array}
    \right] =
    \left[
	\begin{array}{c}
        \hat{t}^* \\
        v^*
	\end{array}
    \right] .
\end{align}
From the definition of the map, we get
\begin{align}
    \dot{f}(\hat{t}^*) &= \frac{k}{2}\frac{1-r}{1+r} , \label{dfts}\\
    v^* &= \frac{k}{2} . 
\end{align}
It is noteworthy that Eq. (\ref{dfts}) has at least one solution for appropriate $\alpha $ because $f(t)$ is $\mathrm{C}^1$-class and periodic.
Then, Jacobian of the linearized equation around the fixed point is 
\begin{align}
    A &= \left[
	\begin{array}{cc}
		1 & 1+r \\
        (1+r)\ddot{f}(\hat{t}^*) & (1+r)^2\ddot{f}(\hat{t}^*)+r^2
	\end{array}
    \right] .
\end{align}
Therefore, if the second-order derivative exists at $\hat{t} = \hat{t}^*$, 
linear stability conditions can be expressed as follows
\begin{align}
    \kappa _c < \ddot{f}(\hat{t}^*) < 0 , \label{lsc}
\end{align}
where the crisis of the stability $\kappa _c$ is defined by
\begin{align}
    \kappa _c \equiv -2\frac{1 + r^2}{(1 + r)^2} . \label{kappa_c}
\end{align}
See Appendix A for detailed calculation.
When $\ddot{f}(\hat{t}^*)=\kappa _c$, the period-doubling bifurcation can exist because the eigenvalue of $A$ with a maximum absolute value is equal to $-1$.

\subsection{$(2, 2k)$-solution}

In the previous subsection, it is suggested that the period-doubling bifurcation may occur regardless of the function form and thus we consider the $(2, 2k)$-solution,
which is the solution generated from the bifurcation of the $(1,k)$-solution.
We set this solution as follows
\begin{align}
    & \left[
	\begin{array}{c}
        \hat{t}_0 \\
        v_0
	\end{array}
    \right] =
    \left[
	\begin{array}{c}
        \hat{t}^*_0 \\
        v^*_0
	\end{array}
    \right] , ~
    \left[
	\begin{array}{c}
        \hat{t}_1 \\
        v_1
	\end{array}
    \right] =
    \left[
	\begin{array}{c}
        \hat{t}^*_1 \\
        v^*_1
	\end{array}
    \right] ,
\end{align}
where $\hat{t}^*_1>\hat{t}^*_0$.
In addition, we require the following condition:
\begin{align}
    \lfloor t^*_1 \rfloor - \lfloor t^*_0 \rfloor = k.
\end{align}
By the representation of the map, we obtain
\begin{gather}
    -\frac{1}{2}(k+\Delta t)^2 + v^*_0(k+\Delta t) + f(\hat{t}^*_0) = f(\hat{t}^*_1), \label{teq1}\\
    -\frac{1}{2}(k-\Delta t)^2 + v^*_1(k-\Delta t) + f(\hat{t}^*_1) = f(\hat{t}^*_0), \label{teq2}\\
    v^*_1 = (1 + r)\dot{f}(\hat{t}^*_1) - r \{v^*_0-(k+\Delta t)\}, \label{veq1} \\
    v^*_0 = (1 + r)\dot{f}(\hat{t}^*_0) -  r \{v^*_1-(k-\Delta t)\} \label{veq2},
\end{gather}
where \(\Delta t = \hat{t}^*_1 - \hat{t}^*_0\).
$v^*_0$ and $v^*_1$ are denoted only by $\hat{t}^*_0$ and $\hat{t}^*_1$.
By taking the sum and the difference of Eqs. (\ref{teq1}) and (\ref{teq2}), 
and substituting $v^*_0$ and $v^*_1$ into them 
under the condition that the order of $f(t)$ is up to three, we obtain
\begin{align}
    & \left(1+\ddot{f}(\hat{t}^*)+\frac{\dddot{f}(\hat{t}^*)}{2}\Delta T\right)\Delta t^2 \notag \\ 
    & ~~~ - 2\dot{f}(t^*)\left\{\ddot{f}(\hat{t}^*)\Delta T + \frac{\dddot{f}(\hat{t}^*)}{4}(\Delta T^2 + \Delta t^2) \right\} = 0 , \label{teq5}
\end{align}
\begin{align}
    & \ddot{f}(\hat{t}^*)-\kappa_c + \frac{\dddot{f}(\hat{t}^*)}{2}\Delta T - \frac{\dot{f}(\hat{t}^*)\dddot{f}(\hat{t}^*)}{3k^2}\Delta t^2 = 0 , \label{teq6}
\end{align}
where $\Delta T = \hat{t}^*_1 + \hat{t}^*_0 - 2t^*$. We assume that $f(t)$ can be expanded in the Taylor series around $\hat{t}^*$.
Refer to Appendix B for detailed calculation.
If $\dddot{f}(\hat{t}^*)=0$, we have
\begin{align}
    \Delta \kappa \equiv \kappa _c - \ddot{f}(\hat{t^*}) = 0.
\end{align}
This means that the $(2, 2k)$-solution exists just at the crisis of the stability of the $(1, k)$-solution if $f(t)$ consists of quadratic functions.
Assuming $\dddot{f}(\hat{t}^*) \neq  0$, we get two equations:
\begin{align}
    & \left(\frac{1+\kappa _c}{\dddot{f}(\hat{t}^*)} - K + \frac{1}{2}\Delta T\right)\Delta t^2 \notag \\ 
    & ~~~ - 2\dot{f}(\hat{t}^*)\left\{\left(\frac{\kappa _c}{\dddot{f}(\hat{t}^*)} - K\right)\Delta T + \frac{1}{4}(\Delta T^2 + \Delta t^2) \right\} = 0 , \label{eqf1}
\end{align}
\begin{align}
    & K + \frac{1}{2}\Delta T - \frac{\dot{f}(\hat{t}^*)}{3k^2}\Delta t^2 = 0 , \label{eqf2}
\end{align}
where $K \equiv  \Delta \kappa /\dddot{f}(\hat{t}^*)$.
For sufficiently small $\dddot{f}(\hat{t}^*)$ with finite $K$, 
Eq. (\ref{eqf1}) can be rewritten as follows
\begin{align}
    & \left(1+\kappa _c\right)\Delta t^2 - 2\dot{f}(\hat{t}^*)\kappa _c \Delta T = 0 . \label{eqf11}
\end{align}
Solving Eqs. (\ref{eqf2}) and (\ref{eqf11}), we obtain
\begin{align}
    \Delta T &= \left(1-\frac{4\dot{f}(\hat{t}^*)^2}{3k^2}\frac{\kappa_c}{1+\kappa_c}\right)^{-1}\frac{\Delta \kappa }{\dddot{f}(\hat{t}^*)}, 
    \label{deltaT} \\
    \Delta t &= \left(\frac{1}{4\dot{f}(\hat{t}^*)}\frac{1+\kappa _c}{\kappa _c}-\frac{\dot{f}(\hat{t}^*)}{3k^2}\right)^{-1/2} \left(\frac{\Delta \kappa }{\dddot{f}(\hat{t}^*)}\right)^{1/2}. \label{deltat}
\end{align}
This approximate solution indicates that 
the bifurcation from the $(1, k)$-solution to the $(2, 2k)$-solution can exist 
even if $\dddot{f}(\hat{t}^*)$ is infinitesimally small.

\section{Numerical result}
In the previous section, we derive the approximate solution in Eqs. (\ref{deltaT}) and (\ref{deltat}) from the necessary conditions for the $(2, 2k)$-solution
when $\dddot{f}(\hat{t}^*)$ and $\Delta \kappa $ are nearly zero.
We numerically demonstrate the existence of the $(2, 2k)$-solution in this section.
We performed the numerical calculation by simulating the motion of the ball with an initial condition which is slightly different from the $(1, k)$-solution.
The solver discretizes time with the time step $10^{-2}$ (the period of the table vibration is $1$) to detect the next collision 
and further applies the bisection method to determine the time of the next collision more accurately.
The maximum error of collision times is $10^{-9}$ unless multiple collisions occur within a single interval of the time step $10^{-2}$ as in a sticking solution \cite{Vogel2011}.
Our solver discretizes time only once in advance, not at each collision time.
In other words, a collision can occur at most once in $10^{-2}$.
This prevents the sticking solution from significantly slowing down the calculations.
It is noteworthy that the $(1, k)$-solution and the $(2, 2k)$-solution can be calculated with $10^{-9}$ accuracy because the intervals of the collision are sufficiently larger than the time step $10^{-2}$.
We run the simulation until the time reaches $5\times 10^{6}$ for each value of $\alpha $, and the data for the final $100$ bounces are plotted in the bifurcation diagrams.
In numerical calculation, we adopt $r = 0.8$, $k = 1$, and
the following function $F_D(t)$ as $F(t)$:
\begin{align}
	F_D(t) \equiv \left\{
	\begin{array}{ll}
		\hat{t}\left(\hat{t}-\frac{1}{2}\right)(D\hat{t}-1) & \left(\hat{t} < \frac{1}{2} \right), \\
        \left(\hat{t}-\frac{1}{2}\right)(\hat{t}-1)(D\hat{t}+1-D) & \left(\hat{t} \ge \frac{1}{2}\right),
	\end{array}
	\right.
\end{align}
where the parameter $D \in [0, 1]$ is introduced to smoothly connect \Okninskiquad \cite{Okninski2012} and \Okninskicubic \cite{Okninski2014}.
$F_D(t)$ consists of two quadratic functions for $D=0$, otherwise it consists of two cubic functions,
as shown in Fig. 2.
\begin{figure}[H]
    \centering
    \includegraphics{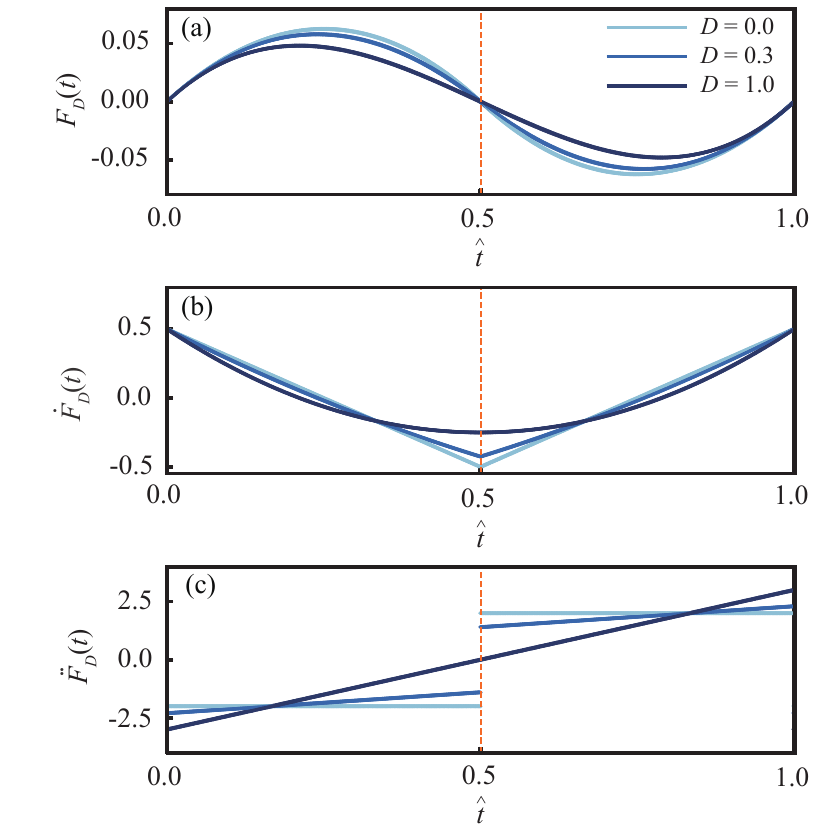}
    \caption{Plots of (a) $F_D(t)$, (b) $\dot{F}_D(t)$, and (c) $\ddot{F}_D(t)$ for representative $D$ values. Dashed lines represent $\hat{t}=1/2$.}
    \label{}
\end{figure}
Our analytical result in Eqs. (\ref{eqf1}) and (\ref{eqf2}) indicates that $(2, 2k)$-solution can exist for non-zero $D$ value.
We plot the bifurcation diagrams for some $F(t)$ in Fig. \ref{bd}.
As shown in Fig. \ref{bd}, the period-doubling bifurcation does not appear for $D=0$, though it is observed in the other cases.
Moreover, the period-doubling bifurcation from the two-period solution to the four-period solution can be seen for $D=1$ case and the sinusoidal case.
The upper row in Fig. \ref{scalingkappa} shows the existence of the period-doubling bifurcation even for the small value of $D$.
Since $\dddot{f}(\hat{t}^*)=6\alpha D$, $D \ll 1$ is equivalent to $\dddot{f}(\hat{t}^*) \ll 1$
and $\Delta \kappa \ll 1$ is satisfied sufficiently close to the bifurcation point.
Therefore, Eqs. (\ref{deltaT}) and (\ref{deltat}) hold near the bifurcation point. Taking the logarithmic forms of both sides yields
\begin{align}
    \log \Delta T &= \log \Delta \kappa -\log \left\{\dddot{f}(\hat{t}^*) \left(1-\frac{4\dot{f}(\hat{t}^*)^2}{3k^2}\frac{\kappa_c}{1+\kappa_c}\right)\right\}  \notag \\
    & \equiv  a_T \log \Delta \kappa  + b_T, \label{logdeltaT} \\
    \log \Delta t &= \frac{1}{2}\log \Delta \kappa -\frac{1}{2}\log \left\{\dddot{f}(\hat{t}^*) \left(\frac{1}{4\dot{f}(\hat{t}^*)}\frac{1+\kappa _c}{\kappa _c}-\frac{\dot{f}(\hat{t}^*)}{3k^2}\right)\right\} \notag \\
    & \equiv  a_t \log \Delta \kappa  + b_t, \label{logdeltat}
\end{align}
where $a_t~(a_T)$ and $b_t~(b_T)$ respectively correspond to the exponent and the coefficient of the power law with respect to $\Delta \kappa$ for $\Delta t~(\Delta T)$ . 
$\dddot{f}(\hat{t}^*) $ depends on not only $D$ but also $\Delta \kappa $.
However, the effect of $\Delta \kappa $ can be ignored when $D$ and $\Delta \kappa $ are sufficiently small.
For more details, see Appendix C.
We plot Eqs. (\ref{logdeltaT}) and (\ref{logdeltat}) and 
estimate $a_T$, $b_T$, $a_t$, and $b_t$ by the least squares method in the lower row of Fig. 4.
It is confirmed that $a_T$ and $a_t$ take values close to $1$ and $1/2$, respectively, for sufficiently small $D$.
Furthermore, we estimate $b_T$ and $b_t$ by setting $a_T=1$ and $a_t=1/2$ as shown in Fig. 5.
Under this condition, we analytically conclude
\begin{align}
    b_T &\sim -\log D - \log\left\{-3\kappa _c\left(1-\frac{4\dot{f}(\hat{t}^*)^2}{3k^2}\frac{\kappa_c}{1+\kappa_c}\right) \right\}, \label{bT} \\
    b_t &\sim -\frac{1}{2}\log D - \frac{1}{2}\log \left\{-3\kappa _c\left(\frac{1}{4\dot{f}(\hat{t}^*)}\frac{1+\kappa _c}{\kappa _c}-\frac{\dot{f}(\hat{t}^*)}{3k^2}\right)\right\} . \label{bt}
\end{align}
Figure 5 suggests that the analytical estimations in Eqs. (\ref{deltaT}) and (\ref{deltat}) give good approximations.

\begin{widetext}
    
\begin{figure}[H]
    \centering
    \includegraphics{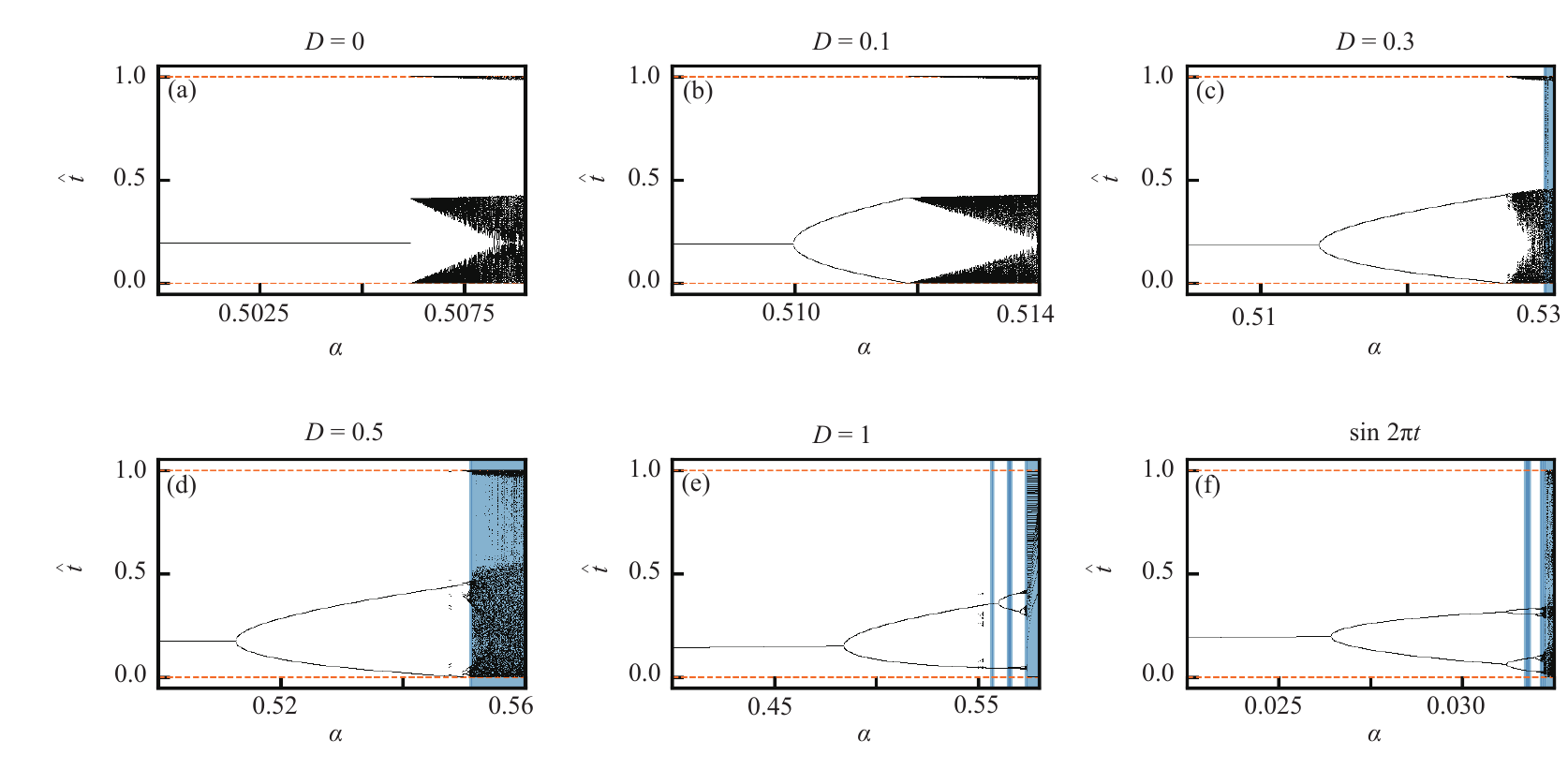}
    \caption{Bifurcation diagrams for 
        (a) $F(t) = F_0(t)$,
        (b) $F(t) = F_{0.1}(t)$,
        (c) $F(t) = F_{0.3}(t)$,
        (d) $F(t) = F_{0.5}(t)$,
        (e) $F(t) = F_1(t)$,
        and (f) $F(t) = \sin 2\pi t$ (for reference).
        Dashed lines represent the domain of $\hat{t}$.
        We used $1000$ values of $\alpha $ in each bifurcation diagram.
        The region where the calculation may be inaccurate to avoid the long calculation time due to sticking solution is filled in blue color.
    }
    \label{bd}
\end{figure}

\begin{figure}[H]
    \centering
    \includegraphics{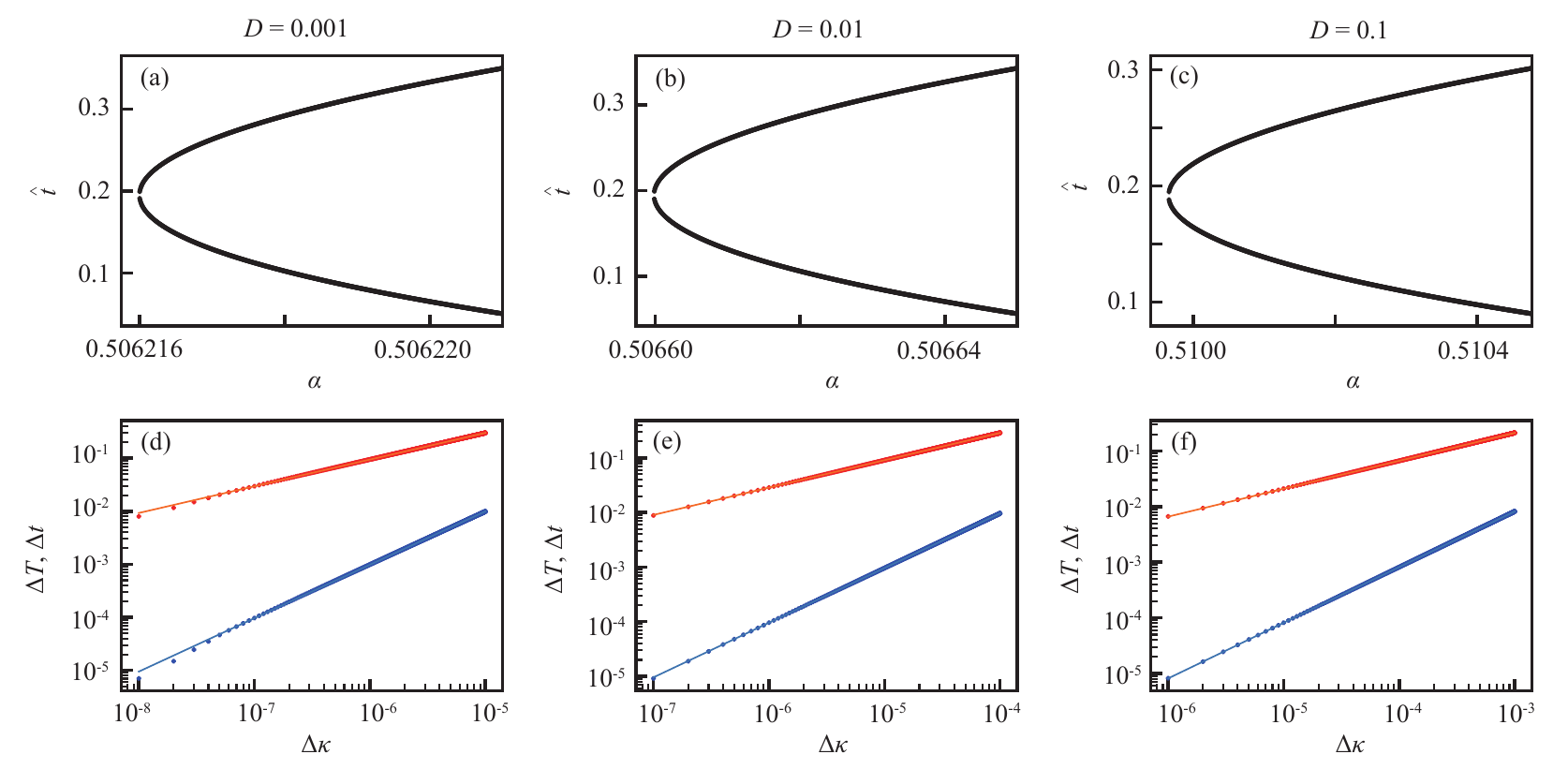}
    \caption{
        Upper row: Enlarged view of the bifurcation diagrams for (a) $D=0.001$, (b) $D=0.01$, and (c) $D=0.1$. 
        We used $1000$ values of $\alpha $ in each bifurcation diagram.
        Lower row: $\Delta T$ (blue) and $\Delta t$ (red) 
        by numerical calculation (points) and least squares fittings (line) for 
        (d) $D=0.001$, (e) $D=0.01$, and (f) $D=0.1$.
        The results of the least squares fittings are
        (d) $\log \Delta T = 1.007 \log \Delta \kappa + 3.031$, 
        $\log \Delta t = 0.503 \log \Delta \kappa + 1.994$, 
        (e) $\log \Delta T = 1.001 \log \Delta \kappa + 1.988$, 
        $\log \Delta t = 0.501 \log \Delta \kappa + 1.462$, 
        and (f) $\log \Delta T = 1.001 \log \Delta \kappa + 0.914$, 
        $\log \Delta t = 0.502 \log \Delta \kappa + 0.830$.
    }
    \label{scalingkappa}
\end{figure}

\end{widetext}

\begin{figure}[htbp]
    \centering
    \includegraphics{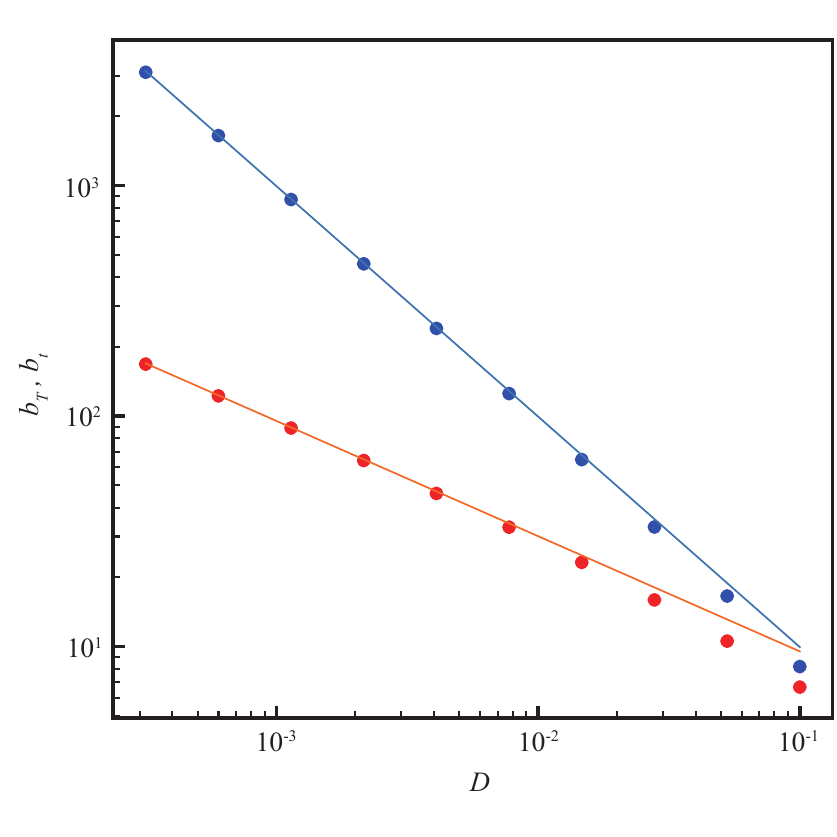}
    \caption{Plots of $b_T$ (blue) and $b_t$ (red) obtained by numerical calculation (points) and analytical results 
    in Eqs. (\ref{bT}) and (\ref{bt}) (lines).}
    \label{scalingD}
\end{figure}

\section{Discussion}
As shown in Fig. \ref{bd}, the route to the chaos seems to be the Feigenbaum scenario \cite{Feigenbaum1978}, 
which is the period-doubling bifurcation sequence to chaos, when $F(t)=\sin 2\pi t$.
This is also true for $F(t)=F_1(t)$, i.e., $D=1$.
On the other hand, the two-period solution disappears when one of the branches corresponding to the two-period solution touches $\hat{t}=0$ and chaos appears immediately afterwards, for $D=0.1$, $D=0.3$, and $D=0.5$. This differs from the Feigenbaum scenario in that the sequence of period-doubling is interrupted by the disappearance of the solution.
The disappearance of the two-period solution in this manner is called border-collision bifurcation \cite{Simpson2009}.
By obtaining the value of $\alpha $ that satisfies $\hat{t}=0$, it may be possible to determine the point where the two-period solution disappears and chaos appears.
In the region of large $D$ values, $\dddot{f}(\hat{t}^*)$ is not sufficiently small and thus the approximate solution in Eqs. (\ref{deltaT}) and (\ref{deltat}) no longer applies.
Therefore, it is necessary to discuss the exact solution for Eqs. (\ref{teq5}) and (\ref{teq6}) to know the point for large $D$ and it may be possible since Eqs. (\ref{teq5}) and (\ref{teq6}) are ascribed to a biquadratic equation about $\Delta t$.
Additionally, one of the branches corresponding to the $2^n$-period ($n\ge 2, n\in \mathbb{N}$) solution may cross $\hat{t}=0$ and the solution may vanish,
though we have not yet confirmed it numerically.

In Fig. 5, the analytical lines deviate from the value obtained by numerical calculation for larger $D$.
There are mainly two reasons for this.
First, the approximate solution in Eqs. (\ref{deltaT}) and (\ref{deltat}) does not hold for the region where $D$ is large.
Second, $\dddot{f}(\hat{t}^*)=6\alpha D$ practically depends on $\Delta \kappa $ because $\alpha $ depends on $D$ and $\Delta \kappa $.
The approximation $\dddot{f}(\hat{t}^*)=-3\kappa _c D$ does not hold for large $D$, as shown in Appendix C.

\section{Conclusion}
We clarified the mechanism of the qualitative differences in bifurcation diagrams between the piecewise quadratic and the piecewise cubic table displacement functions $f(t)$ by focusing on the two-period solution.
We also derived the approximate two-period solution analytically and demonstrated that the approximations explain the numerical result under the assumption of an infinitesimally small cubic term of $f(t)$.

\section*{Acknowledgment}
This work was supported by JSPS KAKENHI Grant Numbers JP19H00749, JP19K14675, JP16H03949. It was also supported by Sumitomo Foundation (No. 181161), the Japan-Poland Research Cooperative Program ``Spatio-temporal patterns of elements driven by self-generated, geometrically constrained flows'' and ``Complex spatio-temporal structures emerging from interacting self-propelled particles" (JPJSBP120204602), and the Cooperative Research Program of ``Network Joint Research Center for Materials and Devices'' (Nos. 20194003 and 20204004).

\appendix
\section{Derivation of the stability of $(1,k)$-solution}
We derive the linear stability condition of the fixed point (\ref{lsc}) in this section.
The trace and the determinant of Jacobian matrix $A$ are
\begin{align}
    \mathrm{tr}\;A &= (1+r)^2\ddot{f}(\hat{t}^*)+1+r^2, \\
    \det A &= r^2.
\end{align}
The eigenvalues $\lambda _\pm $ can be calculated as
\begin{align}
    \lambda _\pm = \frac{\mathrm{tr}\;A \pm \sqrt{(\mathrm{tr}\;A)^2-4\det A}}{2} .
\end{align}
If $(\mathrm{tr}\;A)^2 < 4\det A$, 
\begin{align}
    \max \{|\lambda _\pm |\}= \sqrt{\lambda _+\lambda _-} = r .
\end{align}
Therefore the solution is stable in this case
\begin{align}
    -1 < \ddot{f}(\hat{t}^*) < -\left(\frac{1-r}{1+r}\right)^2 . \label{sc1}
\end{align}
If $(\mathrm{tr}\;A)^2 \ge 4\det A$, 
\begin{align}
    \max \{|\lambda _\pm |\}= \frac{|\mathrm{tr}\;A| + \sqrt{(\mathrm{tr}\;A)^2-4\det A}}{2} . \label{ml2}
\end{align}
From the stability condition that Eq. (\ref{ml2}) is smaller than 1, we get
\begin{align}
    -2\frac{1+r^2}{(1+r)^2} < \ddot{f}(\hat{t}^*) < 0 . \label{sc2}
\end{align}
Considering $0 < r < 1$,
\begin{align}
    -2\frac{1+r^2}{(1+r)^2} < -1, 
\end{align}
and thus the condition
\begin{align}
    -\left(\frac{1-r}{1+r}\right)^2 < 0
\end{align}
is clearly satisfied.
Since Eq. (\ref{sc2}) covers Eq. (\ref{sc1}), the stability condition can be described as follows
\begin{align}
    -2\frac{1+r^2}{(1+r)^2} < \ddot{f}(\hat{t}^*) < 0.
\end{align}

\section{Simplification of the equations for $(2,2k)$-solution}
We derive Eqs.(\ref{teq5}) and (\ref{teq6}) from Eqs. (\ref{teq1})--(\ref{veq2}) in this section.
Taking the sum and the difference, Eqs. (\ref{veq1}) and (\ref{veq2}) yield
\begin{align}	
    & v^*_1+v^*_0 = \dot{f}(\hat{t}^*_1)+\dot{f}(\hat{t}^*_0) + \frac{2r}{1+r}k, \\
    & v^*_1-v^*_0 = \frac{1+r}{1-r}(\dot{f}(\hat{t}^*_1) - \dot{f}(\hat{t}^*_0) ) + \frac{2r}{1-r}\Delta t.
\end{align}
From these, we obtain
\begin{align}	
    & v^*_0 = \frac{1}{2}\left\{ \dot{f}(\hat{t}^*_1)+\dot{f}(\hat{t}^*_0) - \frac{1+r}{1-r}(\dot{f}(\hat{t}^*_1) - \dot{f}(\hat{t}^*_0) )\right\} \notag \\
    & ~~~~~~~~ + \frac{2r}{1+r}k - \frac{2r}{1-r}\Delta t \label{v0}, \\
    & v^*_1 = \frac{1}{2}\left\{ \dot{f}(\hat{t}^*_1)+\dot{f}(\hat{t}^*_0) + \frac{1+r}{1-r}(\dot{f}(\hat{t}^*_1) - \dot{f}(\hat{t}^*_0) )\right\} \notag \\
    & ~~~~~~~~ + \frac{2r}{1+r}k + \frac{2r}{1-r}\Delta t \label{v1}.
\end{align}
Substituting these into the sum and the difference of Eqs. (\ref{teq1}) and (\ref{teq2}), 
\begin{align}
    & \Delta t^2 + (\dot{f}(\hat{t}^*_1)-\dot{f}(\hat{t}^*_0))\Delta t \notag \\ 
    & ~~~ + \frac{1 - r}{1 + r}k\left\{\frac{1 - r}{1 + r}k-(\dot{f}(\hat{t}^*_1)+\dot{f}(\hat{t}^*_0))\right\} = 0 , \label{teq3}\\
    & \left(\dot{f}(\hat{t}^*_1)+\dot{f}(\hat{t}^*_0)-2\frac{1 + r^2}{1 - r^2}k\right)\Delta t \notag \\
    & ~~~ - \frac{1 + r}{1 - r}k(\dot{f}(\hat{t}^*_1)-\dot{f}(\hat{t}^*_0))-2(f(\hat{t}^*_1)-f(\hat{t}^*_0)) = 0 . \label{teq4}
\end{align}
Assuming $f(t)$ as an at-most-third-order function, we performed the Taylor expansion around $\hat{t}^*$ 
of the fixed point as 
\begin{align}
    & \dot{f}(\hat{t}^*_1)-\dot{f}(\hat{t}^*_0) = \ddot{f}(\hat{t}^*)\Delta t + \frac{\dddot{f}(\hat{t}^*)}{2}\Delta t \Delta T , \label{dfm} \\
    & \dot{f}(\hat{t}^*_1)+\dot{f}(\hat{t}^*_0) \notag \\ 
    & ~~~ = 2\dot{f}(\hat{t}^*) + \ddot{f}(\hat{t}^*)\Delta T + \frac{\dddot{f}(\hat{t}^*)}{4}(\Delta T^2+\Delta t^2) , \label{dfp} \\
    & f(\hat{t}^*_1)-f(\hat{t}^*_0) \notag \\
    & ~~~ = \dot{f}(\hat{t}^*)\Delta t + \frac{\ddot{f}(\hat{t}^*)}{2}\Delta t\Delta T + \frac{\dddot{f}(\hat{t}^*)}{24}\Delta t(3\Delta T^2+\Delta t^2) . \label{fm}
\end{align}
Substituting Eqs. (\ref{dfm})--(\ref{fm}) into Eqs. (\ref{teq3}) and (\ref{teq4}) gives
\begin{align}
    & \left(1+\ddot{f}(\hat{t}^*)+\frac{\dddot{f}(\hat{t}^*)}{2}\Delta T\right)\Delta t^2 \notag \\ 
    & ~~~ + \frac{1 - r}{1 + r}k\left\{\frac{1 - r}{1 + r}k-2\dot{f}(\hat{t}^*)-\ddot{f}(\hat{t}^*)\Delta T - \frac{\dddot{f}(\hat{t}^*)}{4}(\Delta T^2 + \Delta t^2) \right\} \notag \\ 
    & ~~~ = 0 , \\
    & \left\{\ddot{f}(\hat{t}^*) + 2\frac{1 + r^2}{(1 + r)^2} + \frac{\dddot{f}(\hat{t}^*)}{2}\Delta T - \frac{1 - r}{1 + r}\frac{1}{k}\frac{\dddot{f}(\hat{t}^*)}{6}\Delta t^2\right\}\Delta t \notag \\
    & ~~~ = 0 .
\end{align}
Using Eq. (\ref{dfts}), we have
\begin{align}
    & \left(1+\ddot{f}(\hat{t}^*)+\frac{\dddot{f}(\hat{t}^*)}{2}\Delta T\right)\Delta t^2 \notag \\ 
    & ~~~ - 2\dot{f}(t^*)\left\{\ddot{f}(\hat{t}^*)\Delta T + \frac{\dddot{f}(\hat{t}^*)}{4}(\Delta T^2 + \Delta t^2) \right\} = 0 , \\
    & \left(\ddot{f}(\hat{t}^*)-\kappa_c + \frac{\dddot{f}(\hat{t}^*)}{2}\Delta T - \frac{\dot{f}(\hat{t}^*)\dddot{f}(\hat{t}^*)}{3k^2}\Delta t^2\right)\Delta t = 0 .
\end{align}
As we focus on the two-period solution, not the fixed point, $\Delta t \neq 0$ is required.
We finally get Eqs. (\ref{teq5}) and (\ref{teq6}).

\section{Approximation of $\dddot{f}(\hat{t}^*)$}
For $f(t) = \alpha F_D(t)$, considering that the fixed point is unstable through period-doubling bifurcation, we obtain
\begin{align}
    \dot{f}(\hat{t}^*) = \alpha \left(3D\hat{t}^{*2}-\left(2+D\right)\hat{t}^*+\cfrac{1}{2}\right) . \label{c1}
\end{align}
From Eq. (\ref{dfts}), $\dot{f}(\hat{t}^*)$ is constant, i.e., independent of $\hat{t}^*$.
Equation (\ref{c1}) is thus a quadratic equation of $\hat{t}^*$, which can be solved as
\begin{align}
    \hat{t}^*_\pm = \frac{(2+D) \pm \sqrt{(2+D)^2-6D\left(1-2\dot{f}(\hat{t}^*)\alpha ^{-1}\right)}}{6D} .
\end{align}
Therefore, $\ddot{f}(\hat{t}^*)$ is described by
\begin{align}
    \ddot{f}(\hat{t}^*_\pm ) &= \alpha \{6Dt^*_\pm-(2+D)\} \notag \\
    &= \pm \alpha \sqrt{(2+D)^2-6D\left(1-2\dot{f}(\hat{t}^*)\alpha ^{-1}\right)} .
\end{align}
From $\ddot{f}(\hat{t}^*)<0$, we have
\begin{align}
    -\alpha \sqrt{(2+D)^2-6D\left(1-2\dot{f}(\hat{t}^*)\alpha ^{-1}\right)} &= \kappa _c+\Delta \kappa .
\end{align}
Squaring both sides and solving the equation with respect to $\alpha $ yield
\begin{align}
    \alpha _\pm &= \frac{1}{D^2-2D+4} \left\{-6D\dot{f}(\hat{t}^*) \right. \notag \\
    &~~~ \left. \pm \sqrt{\left(6D\dot{f}(\hat{t}^*)\right)^2+(D^2-2D+4)(\kappa _c+\Delta \kappa )^2} \right\} .
\end{align}
From $\alpha > 0$, $\alpha = \alpha _+$ is the solution.
Expanding up to the first order of $D$ and $\Delta \kappa $, we finally get
\begin{align}
    \dddot{f}(\hat{t}^*) = 6\alpha D \sim -3\kappa _c D .
\end{align}

\end{document}